%% file: main.tex
\documentclass[prc,aps,floatfix,twocolumn,groupedaddress,nofootinbib,showpacs,preprintnumbers,
amsmath,amssymb,amsfonts,widetable] {revtex4-1}
\usepackage{bm}
\usepackage{soul}
\usepackage{array}
\usepackage{lipsum}
\usepackage{relsize}
\usepackage{mathrsfs}
\usepackage{amssymb}
\usepackage{amsmath}
\usepackage{graphicx}
\usepackage[usenames]{color}
\usepackage{pslatex}
\usepackage{booktabs}
\usepackage{physics}
\usepackage{float}

\newcommand{\Lagrange}[1]{\mathscr{L}_{\raisebox{-0.5pt}{\scriptsize{#1}}}}

\newcommand{\lu}{\resizebox{7pt}{!}{$u$}}

\begin{document}
\title{Role of the isovector spin-orbit potential in mitigating the CREX-PREX dilemma} 

\author{Athul Kunjipurayil and J. Piekarewicz}
\affiliation{Department of Physics, Florida State University, 
Tallahassee, FL 32306, USA}

\author{Marc Salinas}
\affiliation{Lawrence Livermore National Laboratory, 
Livermore, CA 94550, USA}

\date{\today}

\begin{abstract}
Pioneering electroweak measurements of the neutron skin thickness in lead-208 and calcium-48 are 
challenging our understanding of nuclear dynamics. Many theoretical models suggest that the slope of 
the symmetry energy controls the development of a neutron skin in neutron-rich nuclei. This led to the 
expectation that if lead-208 exhibits a large neutron skin, calcium-48 should as well. Given that the PREX 
collaboration reported a relatively thick neutron skin in lead, we anticipated that calcium would also 
have a significant neutron skin. Instead, the CREX collaboration reported a thin neutron skin 
in calcium. Although many suggestions have been proposed, the ``CREX-PREX dilemma" remains 
unsolved. Recently, an intriguing scenario has emerged, suggesting that an enhanced isovector spin-orbit 
interaction could simultaneously account for both results. Following this approach, we performed relativistic 
mean-field calculations with an increased isovector spin-orbit potential. Our findings indicate that while 
this modification significantly affects the structure of calcium-48, it has only a marginal impact on lead-208, 
thereby bringing the results into better agreement with experiment. However, the strong enhancement 
required to mitigate the CREX-PREX dilemma destroys the agreement with a successful spin-orbit 
phenomenology, primarily by modifying the well-known ordering of spin-orbit partners.
\end{abstract}

\maketitle

\section{Introduction}
\label{Sec:Introduction}

The new era of multi-messenger astronomy, which began with the direct detection of gravitational 
waves from the binary coalescence of neutron stars and its associated electromagnetic 
emission\,\cite{Abbott:PRL2017} has created a unique synergy among seemingly distinct fields. 
The contribution from nuclear science to this enterprise is at least twofold. First, nuclear reactions 
under conditions relevant to binary coalescence aim to explain the rapid neutron-capture process, 
which is responsible for the creation of about half of the heavy elements in the 
cosmos\,\cite{Drout:2017ijr,Cowperthwaite:2017dyu,Chornock:2017sdf,Nicholl:2017ahq}. Second, 
the wealth of information acquired from the binary merger and subsequent observations of neutron 
star radii\,\cite{Riley:2019yda,Miller:2019cac,Miller:2021qha,Riley:2021pdl,Choudhury:2024xbk,
Salmi:2024bss} and large masses\,\cite{Cromartie:2019kug,Fonseca:2021wxt}---together with
terrestrial experiments that probe the neutron distribution---have provided critical insights into the 
equation of state (EOS) of neutron-rich matter.

The credit for motivating an experiment that could cleanly determine the neutron distribution of 
atomic nuclei goes to Donnelly, Dubach, and Sick\,\cite{Donnelly:1989qs}. Inspired by the
enormous success in determining the proton distribution using (parity conserving) elastic 
electron scattering, Donnelly, Dubach, and Sick suggested that measuring the parity violating 
asymmetry could provide a model-independent determination of the neutron distribution. 
Although of enormous value as a fundamental nuclear structure quantity, the neutron skin 
thickness---defined as the difference between the neutron and proton rms radii---is known
to serve as a proxy for the pressure of pure neutron matter in the vicinity of nuclear matter 
saturation density\,\cite{Brown:2000,Furnstahl:2001un,RocaMaza:2011pm}. Such unique 
correlation that goes beyond the realm of atomic nuclei, has a significant impact on the 
structure of neutron stars\,\cite{Horowitz:2000xj}, particularly on the radius of low-mass 
neutron stars\,\cite{Horowitz:2001ya,Carriere:2002bx}.

Inspired by the possibility of determining the neutron skin thickness of $^{208}$Pb and its 
impact on neutron-star structure, two different campaigns of the Lead Radius Experiment 
(PREX) were conducted at the Thomas Jefferson National Accelerator Facility. Combining
both experiments the PREX collaboration reported the following value for the neutron skin 
thickness of $^{208}$Pb\,\cite{Abrahamyan:2012gp,Horowitz:2012tj,Adhikari:2021phr}:
\begin{equation}
 R_{\rm skin}^{208}\!=\!R_{n}^{208}\!-\!R_{p}^{208}\!=\!(0.283 \pm 0.071)\,\text{fm}.
 \label{PREX}
\end{equation}
Exploiting the well established correlation between $R_{\rm skin}^{208}$ and the slope 
of the symmetry energy at saturation density $L$\,\cite{Brown:2000,Furnstahl:2001un,
RocaMaza:2011pm}, a large value of $L\!=\!(106 \pm 37)\,\text{MeV}$ was inferred,
suggests that the equation of state of pure neutron matter in the vicinity of nuclear 
matter saturation density is stiff, indicating a rapid increase in pressure with 
density\,\cite{Reed:2021nqk}. The implication for low-mass neutron stars is that the
stellar radius should be large. In turn, given that the tidal deformability scales 
approximately as the fifth power of the neutron star radius\,\cite{Flanagan:2007ix, 
Hinderer:2007mb}, PREX suggests a large tidal deformability.  

Given that a large class of models suggest that the neutron-rich skin of medium to heavy 
nuclei is controlled by the slope of the symmetry energy\,\cite{Piekarewicz:2012pp}, it was 
predicted that the neutron skin thickness of ${}^{48}$Ca would also be large, namely,
$R_{\rm skin}^{\,48}\!=\!(0.229 \pm 0.035)\,{\rm fm}$\,\cite{Piekarewicz:2021jte}. 
Instead, the CREX collaboration extracted a significantly smaller neutron 
skin\,\cite{Adhikari:2022kgg}: 
\begin{equation}
 R_{\rm skin}^{\,48}\!=\!(0.121\pm 0.035)\,\text{fm},
 \label{CREX}
\end{equation}
challenging the assertion that the neutron skins of ${}^{48}$Ca and ${}^{208}$Pb are 
strongly correlated. We note that the quoted CREX error includes both experimental 
and model uncertainties of nearly equal magnitude which have been added in quadrature. 
This is in contrast to the PREX error that is dominated by statistical uncertainties. 
We also note that as a medium-mass nucleus with an interior density that, unlike 
${}^{208}$Pb, does not display a fully saturatated core, ${}^{48}$Ca may not have 
such a profound impact in constraining bulk properties. Nevertheless, both ${}^{48}$Ca 
and ${}^{208}$Pb are instrumental in the validation of theoretical models---particularly 
since ab initio calculations that include three-nucleon forces are now readily 
available\,\cite{Hagen:2015yea}.

Although the theoretical community has devoted a large effort to resolve the CREX-PREX 
``dilemma''\,\cite{Hu:2021trw,Reinhard:2022inh,Zhang:2022bni,Mondal:2022cva,
Papakonstantinou:2022gkt,Yuksel:2022umn,Li:2022okx,Thakur:2022dxb,Miyatsu:2023lki,
Reed:2023cap,Sammarruca:2023mxp,Salinas:2023qic,Roca-Maza:2025vnr}, no convincing 
resolution has been found yet. A key message from these studies is that reconciling the 
large value of $R_{\rm skin}^{208}$ with the constraints set by other nuclear 
observables---now also including the neutron skin thickness of $R_{\rm skin}^{48}$---presents 
an enormous challenge; although see Refs.\cite{Atkinson:2019bwd,Calleya:2024pvj} for a
dispersive-model analysis.

Recently, however, an intriguing proposal has been put forward simultaneously by two 
groups\,\cite{Yue:2024srj,Zhao:2024gjz}. The main tenet espoused by both groups is
the critical role played by the poorly constrained isovector component of the spin-orbit 
force. It is suggested that in order to simultaneously describe both CREX and PREX,
the isovector spin-orbit interaction should be significantly enhanced relative to those
favored in the conventional Skyrme energy density functionals\,\cite{Yue:2024srj}. In
turn, Ref.\cite{Zhao:2024gjz} identifies a delicate interplay between the symmetry 
energy at saturation density and the isovector spin-orbit interaction that may be tuned 
to reproduce the PREX and CREX measurements. An important goal of this latter
reference is to explore the interplay between these quantities in both relativistic
and non-relativistic models. In particular, in the relativistic mean-field models the
isovector spin-orbit interaction is modified by tuning the corresponding isovector
coupling constants after obtaining a Schr\"odinger-like equation by implementing
a Foldy-Wouthuysen transformation. 

In the present paper we also start from the Dirac equation and derive a Schr\"odinger-like 
equation, but without the need for any approximation. That is, the physics content of the 
resulting Schr\"odinger equation is identical to the one from the original Dirac equation. We 
then identify the resulting spin-orbit interaction and modify its isovector component as suggested 
in Refs.\cite{Yue:2024srj,Zhao:2024gjz}. Such a prescription modifies the original correlation 
between the neutron skins of ${}^{48}$Ca and ${}^{208}$Pb, ultimately leading to a much better 
agreement with the experimental results. However, we found that the transition from a largely 
isoscalar to a largely isovector spin orbit interaction required to reproduce the PREX/CREX 
results yields unphysical results, particularly in the ordering of spin-orbit partners---thereby 
destroying the quintessential magic numbers. Moreover, within the scope of the relativistic 
models, it is not possible to modify the spin-orbit interaction without also modifying the central 
potential as they both originate from the same underlying scalar, vector, and tensor interactions. 

The paper has been organized as follows. In Sec.\ref{Sec:Formalism} we develop the formalism
required to obtain a a Schr\"odinger-like equation from the original Dirac equation containing
self-consistent spherically symmetric scalar, vector, and tensor mean fields. Once the resulting
spin-orbit interaction has been derived, we identify the isovector component that will be artificially
enhanced to reproduce the neutron skin thickness of both ${}^{48}$Ca and ${}^{208}$Pb. We 
then proceed to confront in Sec.\ref{Sec:Results} our results against those extracted from the
PREX-CREX analysis. At first glance, we attribute the significant improvement in the description
of the experimental data solely to the modification of the $f_{7/2}$ neutron orbital in ${}^{48}$Ca,
whose wave-function is shifted to the interior, thereby reducing the neutron radius and 
correspondingly the neutron skin thickness\,\cite{Yue:2024srj}. However, in contrast to 
Ref.\cite{Yue:2024srj}, the large enhancement of the isovector spin-orbit interaction leads to
level crossing of spin-orbit partners and the decimation of magic numbers. We offer a summary 
of our results in Sec.\ref{Sec:Summary}.

\section{Formalism}
\label{Sec:Formalism}
\subsection{Covariant Density Functional Theory}

Following the framework developed in a recent publication\,\cite{Salinas:2023qic}, we express 
the underlying Lagrangian density as follows:
\begin{equation}
 \Lagrange{} = \Lagrange{0} + \Lagrange{1} + \Lagrange{2},
\end{equation}
where $\Lagrange{0}$ represent the non-interacting component consisting of the kinetic energy of
all the constituents, which includes the isodoublet nucleon field ($\psi$) and the photon field ($A_{\mu}$)
that accounts for the Coulomb repulsion. In turn, the short-range nuclear interaction is mediated by 
two isoscalar fields---one scalar ($\sigma$) responsible for the intermediate-range attraction and one 
vector ($V_{\mu}$) responsible for the short-range repulsion---and two isovector fields, the scalar
($\boldsymbol{\delta}$) and the vector (${\bf b}_{\mu}$) that account for the isospin dependence 
of the nuclear interaction.

 In turn, $\Lagrange{1}$ represents the Yukawa component of the Lagrangian density which is 
 written entirely in terms of scalar, vector, and tensor bilinears, of both isoscalar and isovector
 character. That is,
\begin{widetext}
\begin{eqnarray}
\Lagrange{1} =
\bar\psi \left[g_{\rm s}\phi   \!+\! g_{{}_{\delta}}{\boldsymbol{\delta}}\!\cdot\!\frac{{\boldsymbol{\tau}}}{2}
             \!-\! \left(g_{\rm v}\gamma^{\;\mu} +  f_{\rm v}\frac{\sigma^{\,\mu\nu}}{2M}\partial_{\nu}\right)\!V_{\mu}
             \!-\! \left(g_{\rho}\gamma^{\;\mu} +  f_{\rho}\frac{\sigma^{\,\mu\nu}}{2M}\partial_{\nu}\right)
                    {\bf b}_{\mu}\!\cdot\!\frac{\boldsymbol{\tau}}{2}
              \!-\! \frac{e}{2}\gamma^{\;\mu}A_{\mu}(1\!+\!\tau_{3})\right]\psi,
 \label{L1}
\end{eqnarray}
\end{widetext}
where we use the standard (Weyl) representation of the Dirac $\gamma$ matrices\,\cite{Peskin1995}, 
$\boldsymbol{\tau}$ is the vector containing the three Pauli matrices, with $\tau_{3}$ being its $z$-component. 
Note that as allowed by Lorentz invariance, the nucleon couples to both $V_{\mu}$ and ${\bf b}_{\mu}$ via vector 
($\gamma^{\;\mu}$) and tensor ($\sigma^{\,\mu\nu}$) couplings. 

Finally, $\Lagrange{2}$ includes both unmixed and mixed meson self-interactions, which have been 
steadily incorporated\,\cite{Boguta:1977xi,Serot:1984ey,Mueller:1996pm,Lalazissis:1996rd,Serot:1997xg,
Horowitz:2000xj,Todd-Rutel:2005fa,Chen:2014sca,Chen:2014mza} since Walecka's original 
work\,\cite{Walecka:1974qa} to refine the relativistic mean field (RMF) models and improve their alignment 
with experimental results. That is, we define $\Lagrange{2}$ as follows:
\begin{widetext}
\begin{equation}
\Lagrange{2} =    - \frac{1}{3!} \kappa\,\Phi^3 - \frac{1}{4!} \lambda\Phi^4 
                            + \frac{1}{4!} \zeta (W_\mu W^\mu)^2 
                            + \frac{1}{4!} \xi (\boldsymbol{B}_\mu \cdot \boldsymbol{B}^{\,\mu})^2  
                            - \Lambda_{\rm s} (\boldsymbol{\Delta} \cdot \boldsymbol{\Delta})\Phi^2
                            + \Lambda_{\rm v} (\boldsymbol{B}_\mu \cdot \boldsymbol{B}^{\,\mu})(W_\mu W^\mu),
\end{equation}
\end{widetext}
where $\Phi\!\equiv\!g_{\rm s}\phi$, $W_\mu\!\equiv\!g_{\rm v}V_\mu$, 
$\boldsymbol{\Delta}\!\equiv\!g_{\delta}\boldsymbol{\delta}$, and 
$\boldsymbol{B}_\mu\!\equiv\!g_{\rho}\boldsymbol{b}_\mu $. The significance of each term may be found in 
Ref.\cite{Salinas:2023qic} and references contained therein.
 
\subsection{Mean Field Approximation}
\label{Sec:MFA}

In the mean-field approximation, the meson-field operators are replaced by their classical expectation values which, 
satisfy non-linear Klein-Gordon equations containing ground-state baryon densities as their source terms\,\cite{Serot:1997xg}. 
In turn, the single-particle orbitals that serve as sources for the Klein-Gordon equations, are obtained from a Dirac Hamiltonian 
that contains scalar, vector, and tensor terms of both isoscalar and isovector character:
\begin{equation} 
 \hat{H} = \bm{\alpha} \cdot \bm{p} + \beta\Big(M - S(r)\Big) + V(r) +i\,\bm{\gamma} \cdot\hat{\bm{r}}\,T(r),
\end{equation} 
where $\bm{\alpha}\!\equiv\!\gamma^{\;0}\bm{\gamma}$, $\beta\!\equiv\!\gamma^{\;0}$ and the scalar, vector, and tensor 
potentials are given as follows:
\begin{subequations}
\begin{align}
S(r) & = \Phi_{0}(r) \pm \frac{1}{2} \Delta_{0}(r), \\
V(r) & = W_{0}(r) \pm \frac{1}{2} B_{0}(r) + eA_{0}(r)\frac{(1\!+\!\tau_{3})}{2},\\
T(r) & = \frac{1}{2M}\left[\frac{f_{\rm v}}{g_{\rm v}}\frac{dW_{0}(r)}{dr}\pm \frac{1}{2} \frac{f_{\rho}}{g_{\rho}}\frac{dB_{0}(r)}{dr}\right],
\end{align}
\label{SVT}
\end{subequations}
where the upper sign is for protons and the lower sign for neutrons. Finally, by invoking spherical symmetry one reduces the Dirac equation 
to the following set of coupled, first-order, ordinary differential equations: 
\begin{subequations}
\begin{eqnarray}
  &&\hspace{-1cm} \left(\frac{d}{dr}+\frac{\kappa^{*}(r)}{r}\right)g_{n\kappa}(r)-\Big(E^{*}(r)+M^{*}(r)\Big)f_{n\kappa}(r)=0,\\
  &&\hspace{-1cm}  \left(\frac{d}{dr}-\frac{\kappa^{*}(r)}{r}\right)f_{n\kappa}(r)+\Big(E^{*}(r)-M^{*}(r)\Big)g_{n\kappa}(r)=0.
\end{eqnarray}
\label{DiracEqns}
\end{subequations}
Here $n$ denotes the principal quantum number (the number of nodes), $\kappa$ is the generalized angular momentum,
and the Dirac potentials are encoded in the following expressions:
\begin{equation}
 M^{*}(r)\!=\!M\!-\!S(r), \;\; E^{*}(r)\!=\!E\!-\!V(r), \;\; \kappa^{*}(r)\!=\!\kappa\!+\!rT(r).
 \label{MEKstar}
\end{equation}

\subsection{The equivalent Schr\"odinger-like equation}
\label{Sec:Schrodinger}

It is the main goal of this section to use the pair of equations displayed in Eqs.(\ref{DiracEqns}) to derive a second-order 
differential equation that for all intents and purposes resembles a Schr\"odinger equation, albeit with energy dependent
potentials. It is important to underscore that the use of ``equivalent" in the title of this section should be taken literally, as
no approximations will be made. Indeed, solutions to the Schr\"odinger equation derived in this section are identical to
those obtained by solving Eqs.(\ref{DiracEqns}).

The first step in the derivation is to uncouple the first order differential equation by expressing the ``small" component
$f_{n\kappa}(r)$ in terms of the ``large" component $g_{n\kappa}(r)$. That is,
\begin{equation}
 f_{n\kappa}(r)= \frac{1}{\Big(E^{*}(r)+M^{*}(r)\Big)}\left(\frac{d}{dr}+\frac{\kappa^{*}(r)}{r}\right)g_{n\kappa}(r),
 \label{fofg}
\end{equation}
which upon substitution into Eq.(\ref{DiracEqns}b) yields the following second-order differential equation:
\begin{widetext}
\begin{equation}
 \left(\frac{d}{dr}\!-\!\frac{\kappa^{*}(r)}{r}\right)
  \!\!\left[\frac{1}{\Big(E^{*}(r)\!+\!M^{*}(r)\Big)}\left(\frac{d}{dr}\!+\!\frac{\kappa^{*}(r)}{r}\right)\right]g_{n\kappa}(r)
 +\Big(E^{*}(r)-M^{*}(r)\Big)g_{n\kappa}(r)=0.
 \label{gofr}
\end{equation}
\end{widetext}
To ensure a clear understanding of the main arguments, we will present the final results and leave the algebraic 
details to the appendix. To start, we introduce the auxiliary wave function $\lu_{n\kappa}(r)$ defined through the 
relation
\begin{equation}
 g_{n\kappa}(r)\!=\!\sqrt{\xi(r)}\,\lu_{n\kappa}(r), \; \text{with}\;\;\xi(r)\!=\!\left(\frac{M^{*}(r)\!+\!E^{*}(r)}{M+E}\right).
 \label{Xi}
\end{equation}
As is customary in the case of the Schr\"odinger equation with spherically symmetric potentials, the re-definition
of the wave function eliminates all first derivatives reducing the problem to an effective one-dimensional 
problem\,\cite{Arnold:1981dt}: 
\begin{equation}
 \left[ \frac{d^{2}}{dr^{2}}-p^{2}-\frac{l(l+1)}{r^{2}}-U_{\rm eff}(r;\kappa,E)\right]\lu_{n\kappa}(r)=0,
 \label{Schrodinger}
\end{equation}
where $p^{2}\!=\!M^{2}-E^{2}$. Although the orbital angular momentum is not conserved, we note that
$l(l+1)\!=\!\kappa(\kappa+1)$, where $l$ is the orbital angular momentum associated to the large component.
This indicates, as in the traditional Schr\"odinger equation, that the third term in the above expression represents
the repulsive centrifugal barrier. In turn, the effective potential contains central, spin-orbit, Darwin\,\cite{Arnold:1981dt},
and tensor terms 
\begin{equation}
 U_{\rm eff}(r;\kappa,E) = U_{\rm c}(r)-(1+\kappa)U_{\rm so}(r)+U_{\rm D}(r)+U_{\rm t}(r),
 \label{Ueff}
\end{equation}
that are defined as follows:
\begin{subequations}
\begin{eqnarray}
  &&\hspace{-1cm}  U_{\rm c}(r) = 2M\left(-S(r)+\frac{E}{M}V(r)\right)+\Big(S^{2}(r)-V^{2}(r)\Big), 
                                \label{Uc} \\
  &&\hspace{-1cm}  U_{\rm so}(r) = -\frac{1}{r}\left(\frac{\xi^{\prime}(r)}{\xi(r)}\right)-\frac{2}{r}T(r), \\
  &&\hspace{-1cm}  U_{\rm D}(r) = \frac{3}{4}\left(\frac{\xi^{\prime}(r)}{\xi(r)}\right)^{2}
                                                     - \frac{1}{2}\left(\frac{\xi^{\prime\prime}(r)}{\xi(r)}\right)
                                                     -\frac{1}{r}\left(\frac{\xi^{\prime}(r)}{\xi(r)}\right), \\
  &&\hspace{-1cm}  U_{\rm t}(r) = T^{2}(r)-T^{\prime}(r)-\frac{2}{r}T(r)+\left(\frac{\xi^{\prime}(r)}{\xi(r)}\right)T(r).
\end{eqnarray}
\label{Potentials}
\end{subequations}
Note that we have suppressed the explicit energy dependence of all the potentials appearing above.
Expressions for $S(r)$, $V(r)$, and $T(r)$ are given in Eqs.(\ref{SVT}) in terms of self-consistent solutions
of the meson fields. In turn, in the absence of a tensor term, the spin-orbit potential may be written as
\begin{equation}
 U_{\rm so}(r) = \frac{1}{r\Big(M^{*}(r)\!+\!E^{*}(r)\Big)} \frac{d}{dr}\Big(S(r)+V(r)\Big).
 \label{Uso0}
\end{equation}
The central potential shown in Eq.(\ref{Uc}), combined with the spin-orbit potential presented in the previous 
equation, represents a hallmark of the covariant framework: (a) a central potential that involves a significant 
cancellation between the relativistic scalar and vector potentials, resulting in moderate binding energies, and 
(b) a large spin-orbit potential, where the scalar and vector potentials add coherently, leading to strong spin-orbit 
splitting, which lies at the core of the nuclear magic numbers\,\cite{Serot:1984ey}. 

As alluded earlier, it is important to underscore that when the self-consistent mean fields given in Eqs.(\ref{SVT})
are used to define the scalar, vector, and tensor potentials---which in turn are used to construct the effective 
Schr\"odinger potential---results for the single particle spectrum are identical as those obtained from directly solving 
the coupled Dirac equation listed in Eq.(\ref{DiracEqns}). Moreover, identical neutron and proton densities to those
obtained from solving the coupled Dirac equation are reproduced by using Eq.(\ref{Xi}) for $g_{n\kappa}(r)$ and 
then generating $f_{n\kappa}(r)$ from Eq.(\ref{fofg}). Both frameworks give identical results.

\section{Results}
\label{Sec:Results}

We presented in the previous section two identical methods to solve the Dirac equation. Whereas solving the 
coupled, first-order Dirac equation is simpler, the Schr\"odinger-like equation provides valuable insights into the 
nature of the dynamics. Moreover, it is only by using the latter method that one can identify---and modify---the
spin-orbit potential. However, unlike the case of the non-relativistic framework, modifying the spin-orbit potential 
is highly model dependent given that all terms appearing in Eqs.(\ref{Potentials}) depend on the scalar and vector
potential, so it is impossible to modify one without modifying the others. Yet, in an attempt to reproduce some of
the results displayed in Refs.\cite{Yue:2024srj,Zhao:2024gjz}, we now proceed to modify the spin orbit potential 
given in Eq.(\ref{Uso0}). In terms of the underlying meson fields listed in Eqs.(\ref{SVT}), the strong-interaction
part (non-Coulomb) part of the spin-orbit potential may be divided into isoscalar and isovector contributions 
as follows:
\begin{equation}
 U_{\rm so}(r) \equiv U^{(0)}_{\rm so}(r) \pm U^{(1)}_{\rm so}(r),
 \label{Uso1}
\end{equation}
where
\begin{align}
 & U^{(0)}_{\rm so}(r) = \frac{1}{r\Big(M^{*}(r)\!+\!E^{*}(r)\Big)} \frac{d}{dr}\Big(\Phi_{0}(r)+W_{0}(r)\Big),\\
 & U^{(1)}_{\rm so}(r) = \frac{1}{2r\Big(M^{*}(r)\!+\!E^{*}(r)\Big)} \frac{d}{dr}\Big(\Delta_{0}(r)+B_{0}(r)\Big).
   \label{Uso2}
\end{align}
In what follows, we examine the impact of artificially enhancing the isovector component of the spin-orbit 
potential\,\cite{Yue:2024srj,Zhao:2024gjz} in mitigating the CREX-PREX dilemma. That is, we let
$U^{(1)}_{\rm so}(r)\!\rightarrow\!\beta\,U^{(1)}_{\rm so}(r)$, with $\beta$ the enhancing parameter.

\begin{center}
\begin{figure}[ht]
\centering
\includegraphics[width=0.4\textwidth]{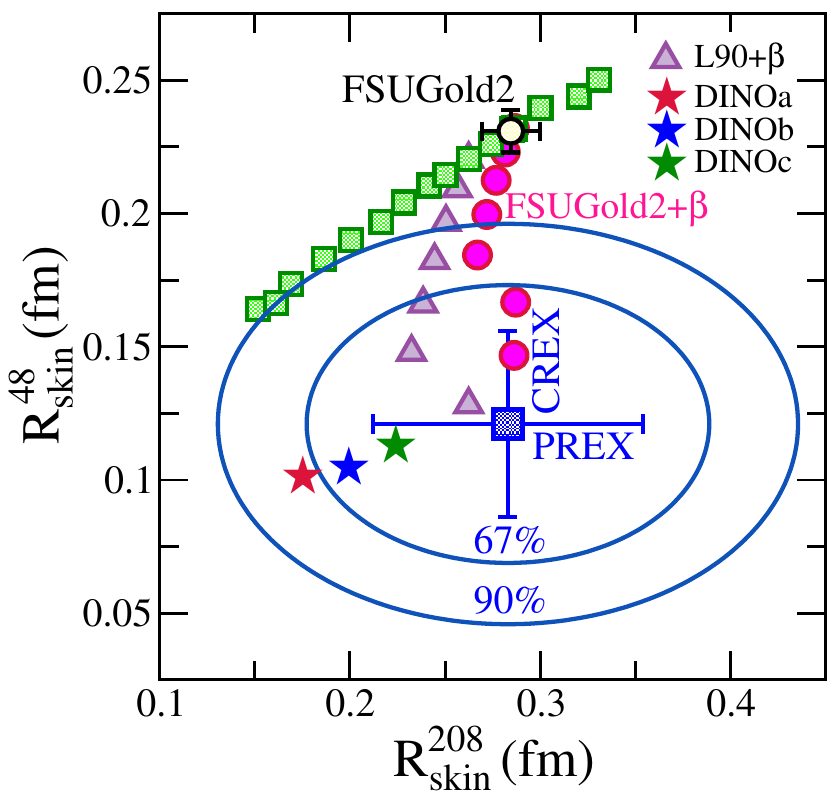}
\caption{Predictions for the neutron skin thickness of ${}^{208}$Pb and ${}^{48}$Ca for a large set of
              covariant RMF models (green squares) alongside the three DINO models introduced in 
              Ref.\,\cite{Reed:2023cap}. The FSUGold2 prediction, denoted with error bars, 
              illustrates typical statistical model uncertainties. In turn, the blue ellipses represent joint 
              PREX and CREX 67\% and 90\% probability contours, as in Fig.\,5 of Ref.\cite{Adhikari:2022kgg}. 
              The evolution with $\beta$ of the predictions from FSUGold2 and FSUGold2(L90) are displayed 
              with circles and triangles, respectively.} 
\label{Fig1}
\end{figure}
\end{center}

We start by displaying in Fig.\ref{Fig1} predictions for the neutron skin thickness of ${}^{208}$Pb and 
${}^{48}$Ca for a large set of covariant RMF models (green squares). These predictions are confronted 
against the PREX-CREX analysis that is displayed as 67\% and 90\% confidence ellipses as in Fig.\,5 
of Ref.\cite{Adhikari:2022kgg}. This set of RMF models clearly encapsulates the CREX-PREX dilemma; 
given that the skin thickness of both nuclei is dominated by the slope of the symmetry energy, a 
``data-to-data" relation emerges: the larger the value of $R_{\rm skin}^{208}$, the larger also the value 
of $R_{\rm skin}^{48}$\,\cite{Piekarewicz:2021jte}. Also shown in the figure are the predictions from the three 
DINO models introduced in Ref.\,\cite{Reed:2023cap}. Whereas these models significantly improve the 
description of the CREX-PREX data, some of the objectionable features of the models---particularly their 
description of ground-state densities---demand additional improvements\,\cite{Salinas:2023qic}.
Lastly, we display in Fig.\ref{Fig1} the predictions from two models---FSUGold2 and FSUGold2(L90)---as 
a function of the the enhancing parameter $\beta$. The covariant FSUGold2 model was calibrated (prior 
to the publication of the neutron skin thickness of lead) assuming a large value 
for $R_{\rm skin}^{208}$---and consequently a large value for the slope of the symmetry energy of 
$L\!=\!(112.8\pm16.1)\,\text{MeV}$\,\cite{Chen:2014sca}. In turn, FSUGold2(L90), with a slightly 
smaller slope of $L\!=\!90\,\text{MeV}$, was derived from FSUGold2 by softening the symmetry energy 
through a fine tuning of the isovector parameters\,\cite{Reed:2021nqk}. Values of $\beta$ explored in
this study are $\beta\!=\!1$ (the original model) and enhancement factors of 
$\beta\!=$\{-50, -100, -150, -200, -250, -300.\} Although these values are large, we note that in most 
of the original RMF models the spin orbit potential is largely isoscalar, so $\beta$ has to be large to 
have any impact on our results. We comment below on the fact that besides being large, the 
enhancement factors are all negative. 

As suggested in Refs.\cite{Yue:2024srj,Zhao:2024gjz}, we observe a dramatic improvement in the 
description of the CREX-PREX data from enhancing the isovector component of the spin-orbit 
potential. Given that the Schr\"odinger-like formalism developed here is free of approximations, the 
predictions from both FSUGold2 and FSUGold2(L90) for $\beta\!=\!1$ fall within the original set of 
models depicted with the green squares. However, the trend with $\beta$ is drastically different.
Indeed, whereas we observe a significant reduction in the neutron skin thickness of ${}^{48}$Ca
the changes to $R_{\rm skin}^{208}$ are fairly moderate. Moreover, for the most extreme value
of $\beta\!=$-300, both predictions fall comfortably within the 67\% confidence ellipse.
\begin{widetext}
\begin{center}
\begin{figure}[ht]
\centering
\includegraphics[width=0.7\textwidth]{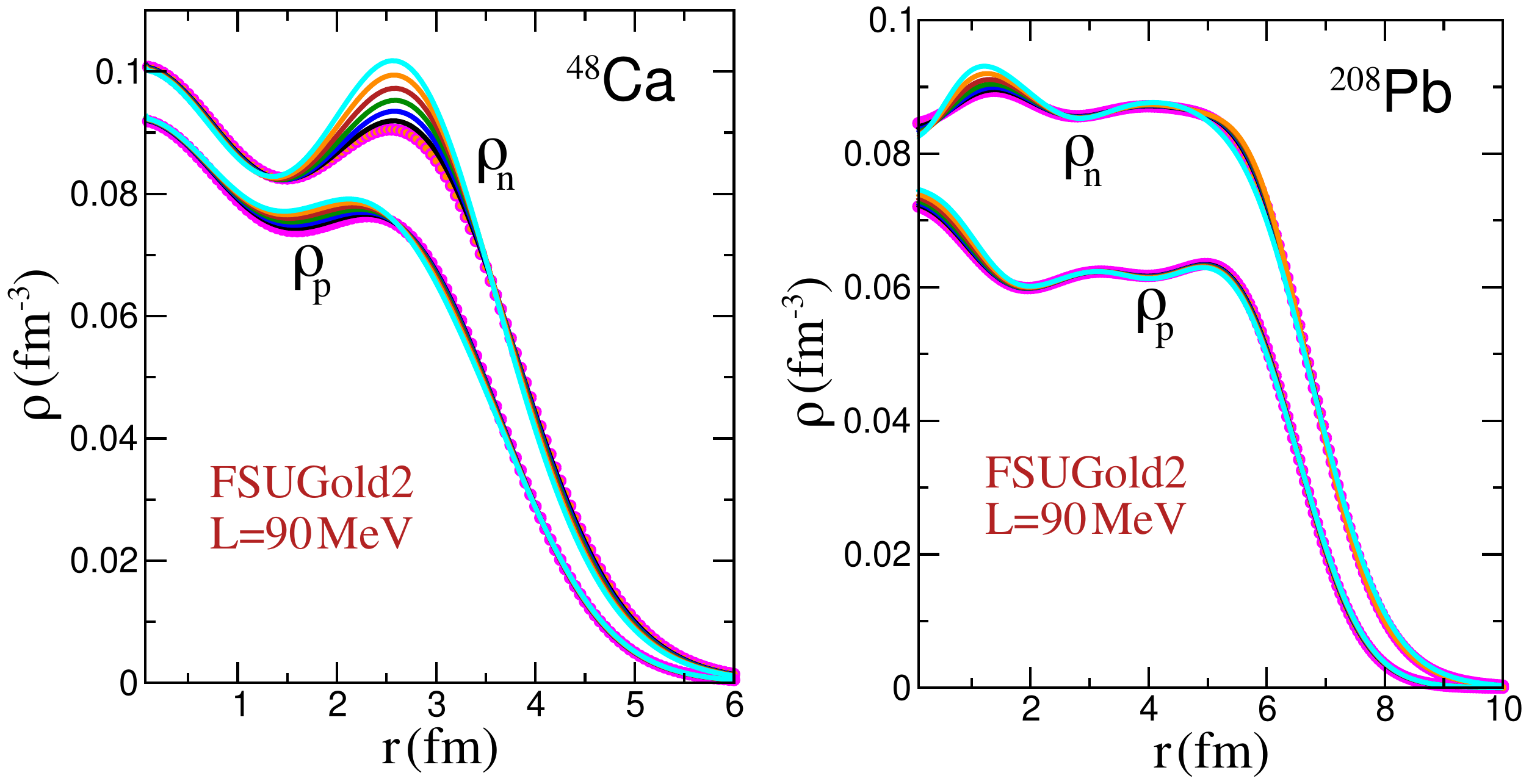}
\caption{Point neutron and proton densities in ${}^{48}$Ca and ${}^{208}$Pb as a function of
the enhancement parameter $\beta$ as predicted by the FSUGold2(L90) model\,\cite{Reed:2021nqk}. 
The magenta line denotes the original ($\beta\!=$1) model and the cyan line the model with an 
enhancement factor of $\beta\!=$-300.}
\label{Fig2}
\end{figure}
\end{center}
\end{widetext}

Although highly encouraging, one must ensure that mitigating the CREX-PREX dilemma
does not come at the expense of spoiling the agreement with other physical observables.
To test the impact of the modification to the isovector spin orbit potential on other physical 
observables, we display in Fig.\ref{Fig2} both neutron and proton densities as a function
of $\beta$ for the FSUGold2(L90) model. We note that the changes to the proton densities
are minimal for both ${}^{48}$Ca and ${}^{208}$Pb. In the case of the neutron density of
${}^{208}$Pb, we observe minor changes that are largely concentrated in the nuclear
interior. As such, this indicates---as verified in Fig.\ref{Fig1}---that modifications to the 
neutron skin thickness of ${}^{208}$Pb due to $\beta$ are minimal. However, large
changes to the neutron density of ${}^{48}$Ca are clearly evident in the figure. 
These changes are concentrated on the surface and the enhancement is almost 
entirely due the valence $f_{7/2}$ neutron orbital. As $\beta$ increases in absolute 
value, the surface bump continues to increase, thereby decreasing the value of the 
neutron radius. This observation is in perfect alignment with the conclusions from 
Ref.\,\cite{Yue:2024srj} that underscore the critical role played by the eight $f_{7/2}$ 
neutrons.

\begin{widetext}
\begin{center}
\begin{figure}[ht]
\centering
\includegraphics[width=0.7\textwidth]{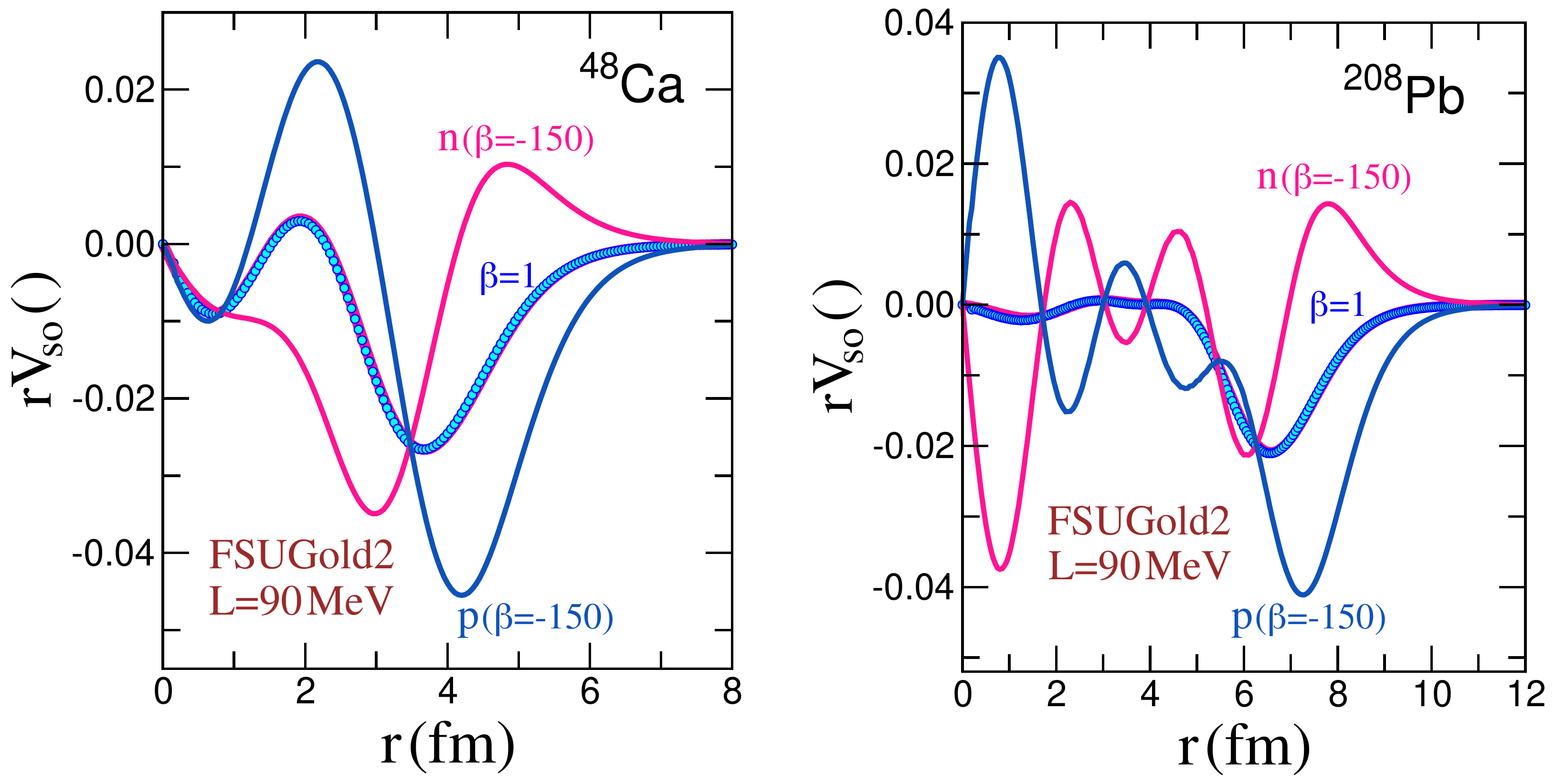}
\caption{Neutron and proton spin orbit potential (multiplied by $r$) as 
              predicted by the FSUGold2(L90) model\,\cite{Reed:2021nqk}.
              Results are shown for the original model ($\beta\!=\!1$) and 
              for one in which the isovector component has been artificially 
              enhanced by a factor of $\beta\!=\!-150$.}
\label{Fig3}
\end{figure}
\end{center}
\end{widetext}

Given the significance of the spin-orbit potential, we now examine its evolution---along with 
that of a selected set of spin-orbit partners---as a function of $\beta$. As we now demonstrate, 
achieving agreement with CREX-PREX data comes at a considerable cost. To maintain the 
readability of Fig.\ref{Fig3}, we only present neutron and proton spin-orbit potentials calculated 
using the FSUGold2(L90) model in two scenarios: the original model with $\beta\!=\!1$ and a 
single modified version with $\beta\!=\!-150$. 

In the original ($\beta\!=\!1$) case, the spin-orbit potential is largely isoscalar, as the neutron 
and proton spin-orbit potentials are practically indistinguishable in the figure. This is expected, 
given that the isoscalar scalar and vector potentials---which add constructively 
[see Eq. (\ref{Uso0})]---are driven by the sum of proton and neutron densities. In contrast, 
the isovector component of the spin-orbit potential is small, as it arises from the difference in 
neutron and proton densities, which in the case of ${}^{48}$Ca is determined almost entirely 
by its eight valence neutrons.

As expected, the picture changes dramatically when large enhancement factors are introduced
to modify the isovector component. Although the magnitude of the spin-orbit interaction changes 
only moderately, the character of the potential shifts drastically---from almost entirely isoscalar to 
predominantly isovector. Interestingly, despite these enormous qualitative changes in the potential 
of both nuclei, the neutron skin thickness of ${}^{208}$Pb remains fairly constant, while that of 
${}^{48}$Ca is significantly reduced (see Fig. \ref{Fig1}). Nevertheless, it appears that such drastic 
modifications to the spin-orbit potential are necessary to reproduce the CREX-PREX data, at least
within the theoretical framework adopted here.

\begin{widetext}
\begin{center}
\begin{figure}[ht]
\centering
\includegraphics[width=0.7\textwidth]{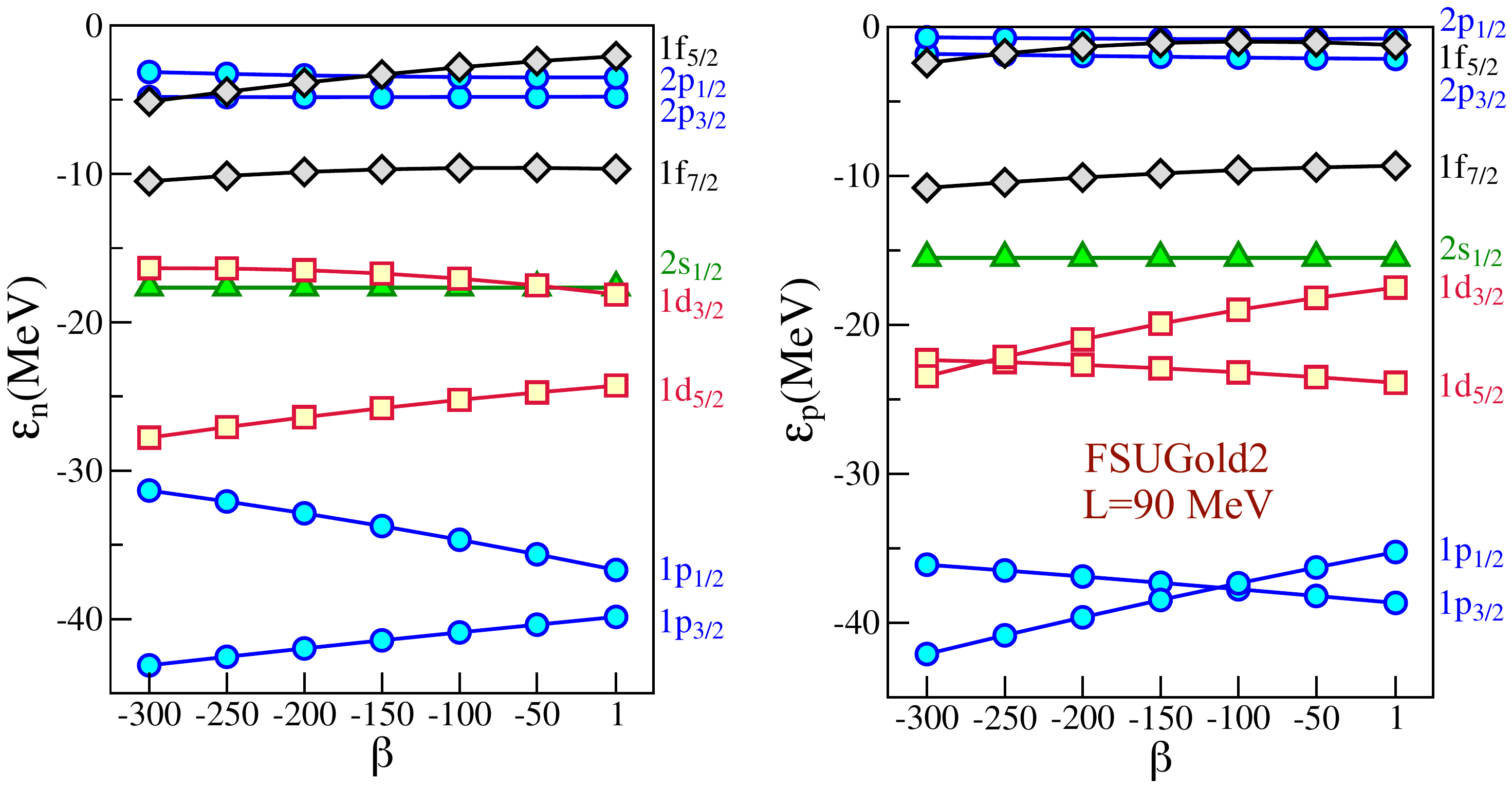}
\caption{Neutron and proton single particle spectra for ${}^{48}$Ca as a function of $\beta$ 
              as predicted by the FSUGold2(L90) model\,\cite{Reed:2021nqk}.}
\label{Fig4}
\end{figure}
\end{center}
\end{widetext}

To fully appreciate the drastic modification of the spin-orbit potential, we present in Fig. \ref{Fig4}
the evolution of the neutron and proton spectra of ${}^{48}$Ca as a function of $\beta$. For
$\beta\!=\!1$, we observe behavior entirely consistent with shell-model phenomenology---namely,
spin-orbit partners with the $j_{\rm max}$ orbital being more strongly bound. In particular, the
strong spin-orbit splitting exhibited by the $f_{7/2}$-$f_{5/2}$ partners is known to be responsible 
for the emergence of the magic number $28$. 

As the isovector component of the spin-orbit potential is artificially enhanced, all neutron orbitals 
with $j\!>\!l$ become more strongly bound. In particular, for the eight $f_{7/2}$neutrons, the 
increase in binding energy---although mild---leads to a smaller neutron radius and, consequently, 
to a reduced neutron skin. However, as $|\beta|$ increases and the isovector component becomes 
dominant, critical features of the successful spin-orbit phenomenology in the proton spectrum are 
lost---most notably, the ordering of the $1p$ and $1d$  spin-orbit partners. Although not explicitly 
shown, the increase in neutron skin thickness of ${}^{208}$Pb for the most extreme values of 
$\beta$ (see Fig. \ref{Fig1}) arises due to the incorrect ordering of the $l\!=\!6$ states. Specifically, 
in these most extreme cases, the $i_{11/2}$ orbital becomes more strongly bound than its 
$i_{13/2}$ spin-orbit partner. Under this unfamiliar scenario, 126 ceases to be a magic number
and ${}^{208}$Pb loses is status as a magic nucleus.

\section{Summary and Outlook}
\label{Sec:Summary}

Given the challenge of simultaneously reproducing the CREX and PREX data, we devoted the 
present paper to examining one of the most recent proposals, namely, the modification of the 
isovector component of the spin-orbit potential\,\cite{Yue:2024srj,Zhao:2024gjz}. To do so in 
the framework of relativistic mean field models requires recasting the Dirac equation in a 
Schrödinger-like form that clearly identifies the various components of the potential. We note 
that the Schrödinger equation derived here was obtained without any approximation. Indeed, 
the physics content of Eq.(\ref{Schrodinger})  is identical to that contained in the original Dirac 
equation. However, we note that unlike non-relativistic descriptions based on Skyrme functionals, 
it is impossible to modify the spin-orbit potential while leaving other parts of the potential unchanged. 
Nevertheless, in an attempt to verify that enhancing the isovector component of the spin-orbit potential 
mitigates the CREX-PREX dilemma, we proceeded along these lines.

In agreement with the results of Ref.\cite{Yue:2024srj} , we found that an enhancement to the 
isovector spin-orbit potential significantly improves the description of the experimental data. 
We also verified that most of the agreement can be attributed to the eight valence neutrons 
occupying the $f_{7/2}$ orbital in ${}^{48}$Ca. Enhancing the spin-orbit interaction increases the 
binding energy of the $f_{7/2}$ orbital, resulting in a reduction of both the neutron radius and the 
neutron skin thickness. While this change is clearly manifested in the neutron density, the proton 
density remains practically unaltered; see Fig.\ref{Fig2}. In the case of ${}^{208}$Pb, neither 
the neutron nor the proton densities were affected. Thus, we found that for models with a stiff 
symmetry energy like the one we used here, the large neutron skin thickness of ${}^{208}$Pb 
is preserved while the corresponding one of ${}^{48}$Ca is reduced.

Although highly encouraging, we decided to examine closely the impact of such a significant
enhancement to the isovector spin-orbit potential. We should note that without the enhancement,
RMF models of the kind explored here favor a spin-orbit potential that is largely isoscalar. The 
reason for the dominance of the isoscalar component is easy to understand: whereas in the 
isoscalar sector the source of the scalar and vector potentials is the sum of neutron and proton
densities, it is the smaller difference in densities that serves as the source of the isovector 
component. In the particular case of of ${}^{48}$Ca, the isovector density is dominated by the eight 
valence neutrons, whereas for the isoscalar case all 48 nucleons contribute to the density. 
However, the large enhancement required to reproduce the CREX-PREX data demands spin\textcolor{red}{-}orbit 
potentials that are significantly different for protons and neutrons;  see Fig.\ref{Fig3}. Unfortunately, 
such large differences spoil the successful shell-model phenomenology. First and foremost, the
well established phenomenology of spin-orbit partners arranged such as the member with the
largest value of $j$ is more strongly bound\textcolor{red}{,} is spoiled. This is clearly observed in the proton
spectrum displayed in Fig.\ref{Fig4} and in the loss of magicity of ${}^{208}$Pb due to the 
crossing of the $l\!=\!6$ orbitals. Hence, one must conclude that although highly suggestive,
the large enhancement of the isovector component of the spin-orbit potential can not be the
solution of the CREX-PREX dilemma, at least within the context of the RMF models explored here. 

So, what's next? In the future, we plan to explore whether a milder enhancement of the isovector 
spin-orbit potential can be accommodated while preserving the successful shell-model phenomenology. 
Achieving this goal requires a multi-pronged approach, culminating in a full Bayesian calibration of 
a new energy density functional. First, adding the delta meson, as in Eq.(\ref{SVT}a), enhances the 
isovector component of the spin-orbit force, as both the delta and rho mesons contribute with the 
same sign. Unfortunately, while the addition of the delta meson helped mitigate the CREX-PREX 
discrepancy, it also introduced large oscillations in the nuclear interior that are not observed 
experimentally\,\cite{Reed:2023cap}. Next, in an effort to exploit the different surface properties of 
${}^{48}$Ca relative to ${}^{208}$Pb, we introduced a tensor coupling that involves derivatives of 
the meson fields; see Eq.(\ref{SVT}c)\,\cite{Salinas:2023qic}. As shown in Eq. (\ref{Potentials}b), 
the resulting tensor potential modifies the spin-orbit interaction. Thus, in principle, one could tune 
the isoscalar tensor coupling to reduce the magnitude of the spin-orbit potential while simultaneously 
adjusting the isovector tensor coupling to enhance the isovector component. 

With the exception of simultaneously reproducing the CREX and PREX data, the class of models used 
in this paper\,\cite{Chen:2014sca,Chen:2014mza} performs well in describing the properties of both finite 
nuclei and neutron stars. In addressing the CREX-PREX discrepancy, however, it is crucial to ensure that 
agreement with experimental and observational data is not compromised. This necessitates a full Bayesian 
calibration containing both additional parameters and experimental data. It is important to note, however, 
that unlike non-relativistic descriptions---where the spin-orbit potential can be adjusted independently of 
other components of the interaction---the Dirac equation inherently includes spin, making such a separation 
impossible. Although enormously challenging, the CREX-PREX dilemma presents a unique opportunity to 
refine our understanding of the nuclear dynamics. After all, one of the main motivations behind the CREX 
campaign was to test and constrain the isovector parts of the energy density functionals. 

\input{./main.bbl}

\begin{acknowledgments}\vspace{-10pt}
This material is based upon work supported by the U.S. Department of Energy Office of Science, Office of Nuclear Physics under Award Number DE-FG02-92ER40750 and under the auspices of the U.S. Department of Energy by Lawrence Livermore National Laboratory under Contract DE-AC52-07NA27344.
\end{acknowledgments} 

\newpage
\onecolumngrid
\section{Appendix}
\label{Sec:Appendix}

As advertised in Sec.\ref{Sec:Schrodinger}, we use this appendix to provide in detail the derivation of the 
Schr\"odinger-like equation that is essential to isolate the various components of the potential. We start 
from Eq.\eqref{gofr}, an expression that already suggests that the upper component $g_{n\kappa}(r)$ of
the Dirac spinor satisfies a second order differential equation. That is,
\begin{equation}
 \left(\frac{d}{dr}\!-\!\frac{\kappa^{*}(r)}{r}\right)
  \!\!\left[\frac{1}{\Big(E^{*}(r)\!+\!M^{*}(r)\Big)}\left(\frac{d}{dr}\!+\!\frac{\kappa^{*}(r)}{r}\right)\right]g_{n\kappa}(r)
 +\Big(E^{*}(r)-M^{*}(r)\Big)g_{n\kappa}(r)=0.
 \label{gofr}
\end{equation}
We now proceed to rewrite the above equation in a form that enables one to separate the various terms
that will need to be manipulated. In what follows and for simplicity, we suppress the quantum numbers 
``$n\kappa$''. We then write
\begin{equation}
  \!\!\!\frac{d}{dr}\!\!\left[\frac{1}{\Big(E^{*}(r)\!+\!M^{*}(r)\Big)}\left(g'(r)+\frac{\kappa^{*}(r)}{r}g(r)\right)\right]
  \!-\!\frac{\kappa^{*}(r)}{r}\!\!\left[\frac{1}{\Big(E^{*}(r)\!+\!M^{*}(r)\Big)}\left(g'(r)\!+\!\frac{\kappa^{*}(r)}{r}g(r)\right)\right]
 \!=\!-\!\Big(E^{*}(r)-M^{*}(r)\Big)g(r).
 \label{gofr1}
\end{equation}
We now manipulate each term appearing on the left-hand side of the equation separately. For the first term one obtains
\begin{equation}
\begin{split}
 \!\!\!\frac{d}{dr}\!\!\left[\frac{1}{E^{*}(r)\!+\!M^{*}(r)}\left(g'(r)+\frac{\kappa^{*}(r)}{r}g(r)\right)\right] 
    = & \frac{1}{M^{*}(r)+ E^{*}(r)} 
    \left[-\frac{\xi'(r)}{\xi(r)}\left(g'(r)+\frac{\kappa^{*}(r)}{r}g(r)\right)\right] \\
   + & \frac{1}{M^{*}(r)+ E^{*}(r)} 
    \left[g''(r)+\frac{\kappa^{*}(r)}{r}g'(r) - \frac{\kappa^{*}(r)}{r^2}g(r) + \frac{\kappa^{*'}(r)}{r}g(r)\right],
\end{split}
\end{equation}
where $\xi(r)$ was already defined in Eq.\eqref{Xi} as 
\begin{equation}
 \xi(r)\!=\!\left(\frac{M^{*}(r)\!+\!E^{*}(r)}{M+E}\right).
\end{equation}
In turn, the second term on the left-hand side of Eq.\eqref{gofr1} may be simply written as follows:
\begin{equation}
    -\frac{\kappa^{*}(r)}{r}\!\!\left[\frac{1}{\Big(E^{*}(r)\!+\!M^{*}(r)\Big)}\left(g'(r)\!+\!\frac{\kappa^{*}(r)}{r}g(r)\right)\right]
    = \frac{1}{M^{*}(r)+ E^{*}(r)} \left[-\frac{\kappa^{*}(r)}{r}g'(r)-\frac{\kappa^{*2}(r)}{r^{2}}g(r)\right].  
\end{equation}
Using these expanded expressions one can now rewrite Eq.\eqref{gofr} in the following suggestive form:
\begin{equation}
   \left[g''(r)- \frac{\xi'(r)}{\xi(r)}g'(r) - \frac{\xi'(r)}{\xi(r)}\frac{\kappa^{*}(r)}{r}g(r)
   -\frac{\kappa^{*}(r)\Big(\kappa^{*}(r)+1\Big)}{r^2}g(r) -\frac{\kappa^{*'}(r)}{r}g(r) \right] 
   = \Big(M^{*2}(r)- E^{*2}(r)\Big)g(r).
   \label{g''}
\end{equation}
The last step required to conform with the traditional one-dimensional form of Schr\"odinger's equation displayed
in Eq.\eqref{Schrodinger} is to eliminate the term containing a first derivative in the above expression. To do so
we express $g(r)$ as follows:
\begin{equation}
    g(r) = \alpha(r) u(r),
    \label{galpha}
\end{equation}
where the auxiliary function $\alpha(r)$ has been introduced with the sole purpose of eliminating terms containing
first derivatives in $u(r)$. This can be accomplished by selecting $\alpha(r)$ to satisfy the following first-order 
differential equation [see Eq.\eqref{Xi}]:
\begin{equation}
 \alpha'(r) - \frac{1}{2}\frac{\xi'(r)}{\xi(r)} \alpha(r) =0 \; \rightarrow\; \alpha(r)=\sqrt{\xi(r)}.
\label{alpha}
\end{equation}
Using both Eqs.\eqref{galpha} and \eqref{alpha} to compute the first and second derivatives of $g(r)$, which are
then substituted in Eq.\eqref{g''}, yields the desired form of the Schr\"odinger equation. That is,
\begin{equation}
 \left[ \frac{d^{2}}{dr^{2}}-p^{2}-\frac{l(l+1)}{r^{2}}-U_{\rm eff}(r;\kappa,E)\right]\lu_{n\kappa}(r)=0,
\end{equation}
where the effective potential $U_{\rm eff}(r;\kappa,E)$ is given in Eqs.(\ref{Ueff}-\ref{Potentials}) 
of Sec.\ref{Sec:Schrodinger}. Note that in obtaining the above equation we defined $p^{2}\!=\!M^{2}-E^{2}$ 
and used the relation $\kappa(\kappa+1)\!=l(l+1)$.
\vfill\eject

\end{document}

%% file: main.bbl
%